\documentclass[aps,prl,superscriptaddress,twocolumn,no-footinbib,showpacs,floatfix,amsmath,amssymb]{revtex4-1}

\usepackage{graphicx}% Include figure files
\usepackage{dcolumn}% Align table columns on decimal point
\usepackage{bm}% bold math
\usepackage[normalem]{ulem}
\usepackage{subfigure} 
\usepackage{soul,xcolor}
\usepackage{color}
\usepackage{siunitx}
\usepackage{amsmath}
\usepackage{amsfonts}
\usepackage{amssymb}
\usepackage{textcomp}
\setstcolor{red}

\RequirePackage{color}
\usepackage[utf8]{inputenc}
\usepackage[T1]{fontenc}
\usepackage{mathptmx}

    \newcommand{\angstrom}{\mbox{\normalfont\AA}}

	\renewcommand{\vec}[1]{  \boldsymbol{\mathbf{#1}}  }

	\newcommand{\kk}{\mathbf{k}}
	\newcommand{\qq}{\mathbf{q}}
	
	\newcommand{\GG}{\mathbf{G}}
	\newcommand{\KK}{\mathbf{K}}
	
	\newcommand{\RR}{\mathbf{R}}
	\newcommand{\ddelta}{\boldsymbol{\delta}}

	\newcommand{\rr}{\mathbf{r}}

	\newcommand{\re}{\mathrm{Re}}
	\newcommand{\im}{\mathrm{Im}}
\begin{document}

\title{Multi-flavor Dirac fermions in Kekul\'{e}-distorted graphene bilayers}

\author{David A.\ Ruiz-Tijerina}
\email{david.ruiz-tijerina@cnyn.unam.mx}
\affiliation{Centro de Nanociencias y Nanotecnolog\'ia, Universidad Nacional Aut\'onoma de M\'exico. Apdo.\ postal 14, 22800, Ensenada, Baja California, M\'exico}

\author{Elias Andrade}
\affiliation{
 Depto. de Sistemas Complejos, Instituto de F\'{i}sica, Universidad Nacional Aut\'{o}noma de M\'{e}xico (UNAM). Apdo. Postal 20-364, 01000 M\'{e}xico D.F., M\'{e}xico
 }%

\author{Ramon Carrillo-Bastos}
%\email{ramoncarrillo@uabc.edu.mx}
\affiliation{Facultad  de  Ciencias,  Universidad  Aut\'onoma  de  Baja  California, 22800  Ensenada,  Baja  California,  M\'exico.}

\author{Francisco Mireles}
\affiliation{Centro de Nanociencias y Nanotecnolog\'ia, Universidad Nacional Aut\'onoma de M\'exico. Apdo.\ postal 14, 22800, Ensenada, Baja California, M\'exico}
%\email{fmireles@cnyn.unam.mx}

\author{Gerardo G. Naumis}
\email{naumis@fisica.unam.mx}
%\homepage{\\http://www.fisica.unam.mx/personales/naumis/}
\affiliation{
 Depto. de Sistemas Complejos, Instituto de F\'{i}sica, Universidad Nacional Aut\'{o}noma de M\'{e}xico (UNAM). Apdo. Postal 20-364, 01000 M\'{e}xico D.F., M\'{e}xico
 }%
\date{\today}
%\maketitle

\begin{abstract}
Graphene's electronic structure can be fundamentally altered when a substrate- or adatom-induced Kekul\'e superlattice couples the valley and isospin degrees of freedom. Here, we show that the band structure of Kekul\'e-textured graphene can be re-engineered through layer stacking. We predict a family of Kekul\'e graphene bilayers that exhibit band structures with up to six valleys, and room-temperature Dirac quasiparticles whose masses can be tuned electrostatically. Fermi velocities half as large as in pristine graphene put this system in the strongly coupled regime, where correlated ground states can be expected. 
\end{abstract}

%\begin{abstract}
%Graphene is widely known for its low-energy electronic excitations, consisting of massless Dirac fermions with opposite chiralities, located at inequivalent Brillouin zone corners. This is fundamentally altered by a substrate- or adatom-induced Kekul\'e superlattice, which couples both fermion species through Bragg scattering, and predicted to either open a much-sought-after mass gap or introducing a new valley-isospin chiral symmetry. Here, we show that, analogously to pristine graphene, the band structure of Kekul\'e-textured graphene can be re-engineered through layer stacking. We propose a family of Y-shaped Kekul\'e graphene bilayers, with distinct symmetry properties that may be approximated by the wallpaper groups $p6m$, $cmm$, $p6$, and $cm$. All structures discussed exhibit multi-valley band structures, with Dirac-like quasiparticles whose masses can be tuned electrostatically, by means of out-of-plane electric fields. In particular, the highly symmetric $p6m$ structure features an unprecedented six flavors of spin-degenerate, anisotropic Dirac fermions. Fermi velocities six times smaller than in pristine graphene put this system in the strongly coupled regime, where correlated ground states can be expected.
%\end{abstract}

\maketitle

Superlattices naturally emerge in multilayered two-dimensional (2D) crystals, either by differences in lattice constants or layer orientation. Heterostructures formed by stacking nearly commensurate van der Waals crystals\cite{GeimNature2013} display moir\'e patterns, with approximate periodicities tens of times larger than the materials' lattice constants. This long-range periodicity has enabled the observation of Hofstadter's butterfly in graphene on hexagonal boron nitride (hBN)\cite{hunt_GhBN_2013,dean_GhBN_2013}, and more recently of exciton minibands\cite{twist_angle2018,tutuc,XiaodongXu,Chenhao} and moir\'e-umklapp optical signatures\cite{twist_angle2018} in hetero-bilayers of transition-metal dichalcogenides. Moir\'e patterns also form in twisted graphene bilayers, where the conduction and valence bands flatten at so-called ``magic angles'' \cite{macdonald_pnas}, leading to strongly correlated phenomena such as unconventional superconductivity and many-body insulating behavior\cite{JH_SC,JH_Mott}.

A different type of superlattice, caused by Kekul\'e distorsions, has also been theorized\cite{Chamon2000,cheianov_2009} and  recently reported in graphene monolayers on copper substrates at room temperature\cite{gutierrez}. In the latter case, so-called ghost copper adatoms distributed periodically beneath the graphene plane produce a bond density wave that triples the graphene unit cell over nanometre-sized regions, where charge carriers comprise two different species of valley-pseudospin-locked Dirac fermions\cite{gamayun,Eliaspaper}. One may envision Kekul\'e-distorted graphene as a new building block for multilayered 2D materials, with new electronic properties that can be engineered by controlling the stacking type. However, whether the substrate effect causing the Kekul\'e distortion will propagate to higher layers is, to our knowledge, still an open question.

\begin{figure}[t!]
    \centering
    \includegraphics[width=0.8\columnwidth]{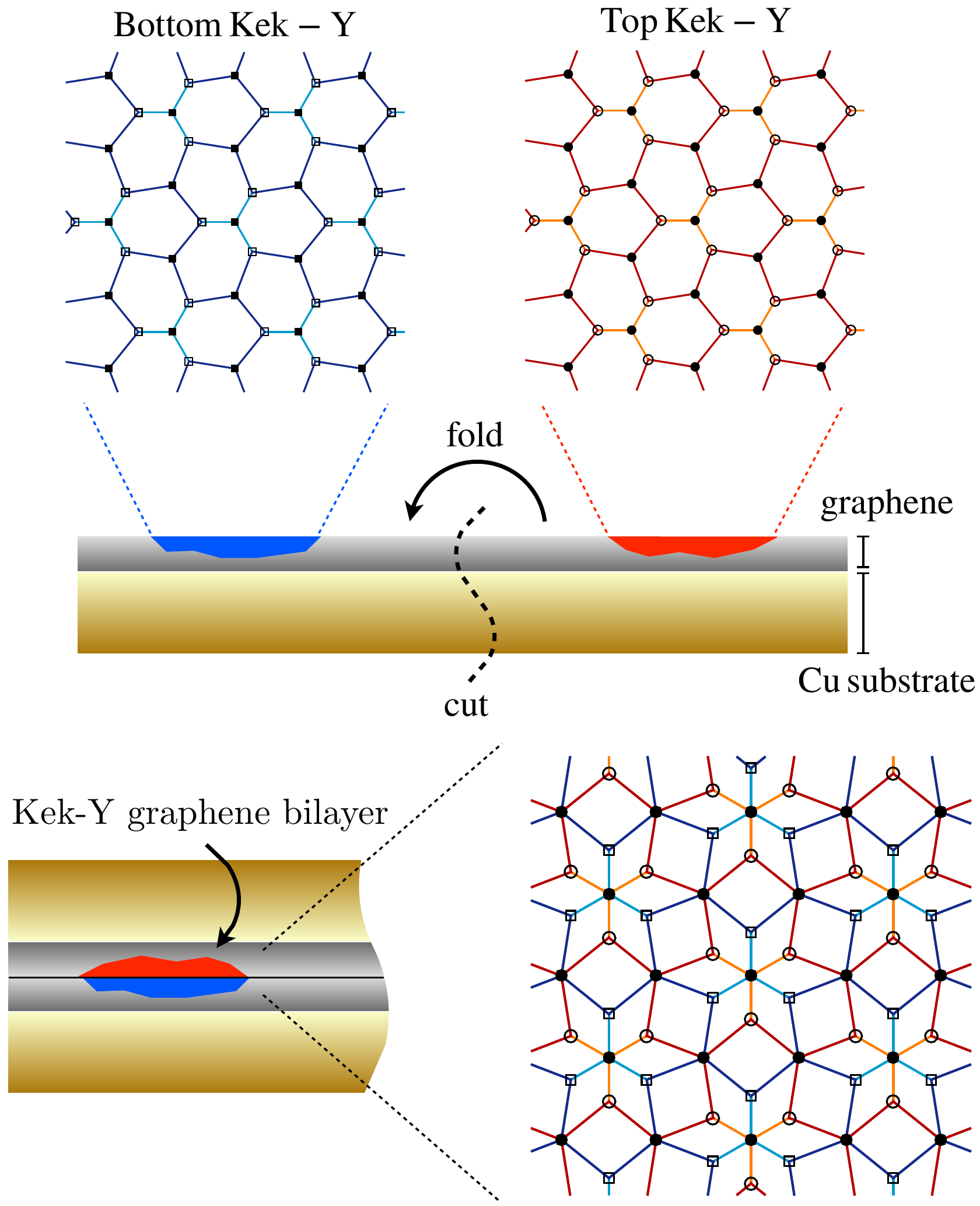}
    \caption{Top: Epitaxial graphene on a copper substrate, showing two extended Kekul\'e-distorted regions (exaggerated for clarity), highlighted in blue and red. Bottom: Cutting and folding the sample, a Bernal-stacked bilayer region can be formed, where the two graphene layers have Kek-Y deformations with opposite orientation.}
    \label{fig:device}
\end{figure}

In this Letter, we show theoretically that Kekul\'e-distorted graphene bilayers support up to six flavors of valley-degenerate Dirac quasiparticles, with electrostatically tunable relativistic masses. While these low-energy Dirac bands are directly related to the band topology transition below energies of order $1\,{\rm meV}$ in bilayer graphene\cite{mccann_falko}, we predict band widths of order $20\,{\rm meV}$, preserving multi-valley physics almost at room temperature, and for carrier dopings of up to $10^{12}{\rm cm}^{-2}$.

We propose a system where two graphene layers with Kekul\'e distortions are stacked with $AB$ (Bernal) configuration, for instance, following the procedure depicted in Fig.\ \ref{fig:device}. Such a system may be produced by combining the fabrication of Kekul\'e-textured graphene on copper samples\cite{gutierrez} with state-of-the-art manual transfer techniques used to fabricate van der Waals heterostructures\cite{JH_SC,JH_Mott,twist_angle2018,tutuc,XiaodongXu,Chenhao}. Without relative rotation of the layers, and by considering the possible Kekul\'e-center stacking types, we predict a family of $5$ patterns that may be catalogued based on their approximate 2D symmetries by the wallpaper groups $p6m$, $cmm$, $p6$, and $cm$ (Supplement). This allows for a rich playground to explore quantum phases.

Among these, we consider the low-energy bands of bilayer superlattices with  $p6m$ wallpaper symmetry, which exhibit a multi-valley band structure with six flavors of massless Dirac fermions. The corresponding states are delocalized across the bilayer, such that a mass gap can be opened and tuned by out-of-plane electric fields, by contrast to Dirac fermions in monolayer graphene.  The Dirac quasiparticles in the $p6m$ Kekul\'e bilayer are six times slower than those in single graphene layers, and thus much more susceptible to many-body phenomena driven by Coulomb interactions, such as spontaneous symmetry breaking\cite{lemonik2010,semenoff2012} and many-body insulating phases\cite{MacDonald2012}.

%These cones appear as the nodes produced by the Lifshitz transition at both valleys of bilayer graphene\cite{mccann_falko} are folded onto the center of the Brillouin zone, and mixed by the periodic bond distortion. Nonetheless, the Dirac cones appear at a new momentum scale, introduced by the Kekul\'e distortion strength.

\emph{Model.} The envisioned system may be fabricated from a single layer of epitaxial graphene grown on a copper substrate, cut and folded such that two regions with Y-shaped Kekul\'e distortions (Kek-Y)\cite{gutierrez} become stacked, forming interlayer bonds. One of the possible outcomes, which will be the focus of this Letter, is sketched in Fig.\ \ref{fig:device}, where the stacking is $AB$ (Bernal) type, and the Kek-Y distortion centers of both layers coincide. Due to the folding, the two Kek-Y distorsions are oriented with a relative angle of $180^\circ$, yielding a structure whose in-plane symmetry is described by the $p6m$ wallpaper group. In three dimensions, this configuration is centrosymmetric, resulting in a spin-degenerate band structure in the absence of external magetic fields. Thus, neglecting spin-orbit interactions induced by the substrate, a spinless Hamiltonian will be considered in the following.

\begin{figure*}[t!]
\begin{center}
\includegraphics[width=2.0\columnwidth]{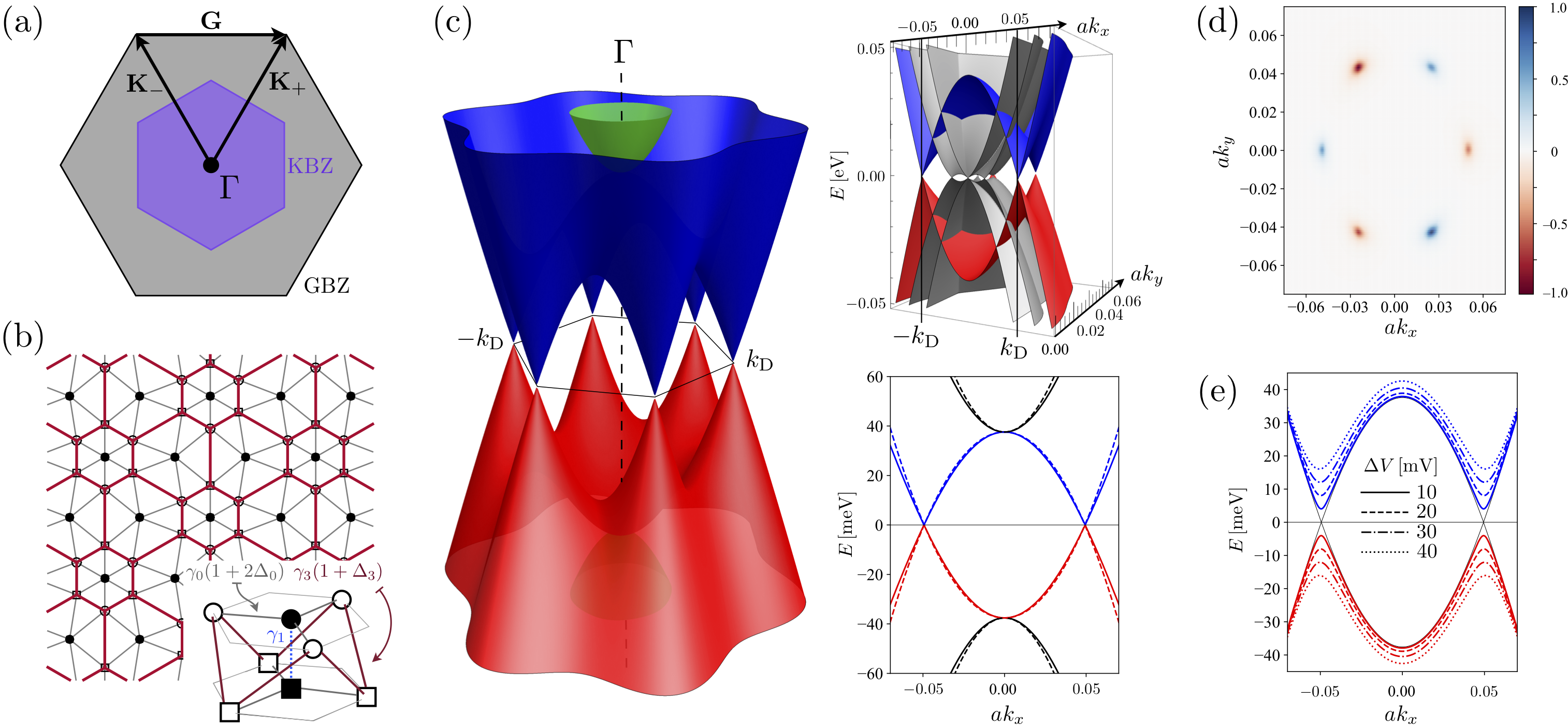}
\caption{(a) Pristine (GBZ) and Kekul\'e-distorted (KBZ) graphene Brillouin zones. (b) Kek-O pattern formed by the bonds between top-layer $A$ and bottom-layer $B$ sublattice atoms when projected onto the plane, in the stacking configuration of Fig.\ \ref{fig:device}. The inset shows a sketch of the central hexagon, in three dimensions. (c) Multi-valley band structure of the symmetric Kek-Y graphene bilayer of Fig.\ \ref{fig:device}, consisting of two inequivalent sets of Dirac bands, with opposite chiralities $\eta = \pm 1$. Top inset: comparison between the multi-valley structure in Kek-Y bilayer graphene, and the multi-node structure found in pristine bilayer graphene, below the Lifshitz transition, when folded into the KBZ (gray). Bottom inset: comparison between the Dirac bands of $H_{\rm eff}$ (dashed lines) and the full Hamiltonian (solid lines), along the $k_y=0$ line of the KBZ. (d) Berry curvature (arbitrary units) near the Dirac points, considering a small mass term to avoid divergences. (e) Application of an out-of-plane electric field causes a potential drop $\Delta V$ across the bilayer, giving a finite mass to the Dirac fermions. All calculations performed with $\Delta_0=0.08$.}
\label{fig:main}
\end{center}
\end{figure*}
We model the graphene monolayers as%\footnote{We ignore on-site energy terms associated to (i) intra-layer sublattice asymmetry and (ii) molecular field asymmetry between dimer and non-dimer sites. (i) produces only a slight electron-hole asymmetry, with no qualitative effects to our results, whereas (ii) has similar effects to an out-of-plane electric field (see Fig.\ \ref{fig:main}e). Both terms have been measured\cite{Basov2009}, with values $\sim 10\,{\rm meV}$.}
\begin{equation}\label{eq:H0}
	H_0=\sum_{\RR}\sum_{j=1}^3\left[ t_{\rm B}(\RR,\ddelta_j)a_{\RR}^\dagger b_{\RR+\ddelta_j} + t_{\rm T}(\RR-\ddelta_j,\ddelta_j)c_{\RR-\ddelta_j}^\dagger d_{\RR} + \text{H.c.} \right],
\end{equation}
where the annihilation operators $a_{\rr}$ and $b_{\rr}$ ($c_{\rr}$ and $d_{\rr}$) correspond to the $A$ and $B$ sublattices of the bottom (top) layer, respectively; $\ddelta_j$ are the nearest-neighbor vectors $\ddelta_1=\tfrac{a}{\sqrt{3}}(\tfrac{\sqrt{3}}{2},\,\tfrac{1}{2})$, $\ddelta_2=\tfrac{a}{\sqrt{3}}(-\tfrac{\sqrt{3}}{2},\,\tfrac{1}{2})$ and $\ddelta_3=\tfrac{a}{\sqrt{3}}(1,\,0)$, with $a=2.461\angstrom$ the lattice constant; and $\RR=m\vec{a}_1+n\vec{a}_2$ ($\vec{a}_1=\ddelta_3-\ddelta_1$, $\vec{a}_2=\ddelta_3-\ddelta_2$; $m$ and $n$ integers) are the bottom-layer $A$ sublattice sites. In writing Eq.\ \eqref{eq:H0}, we ignore intralayer sublattice asymmetries; the role of other on-site energy asymmetries in breaking the $p6m$ symmetry will be discussed below\cite{Marcin2010}.

In Bernal stacking, the top-layer lattice can be obtained from the bottom one, by an in-plane displacement of $-\ddelta_3$, and an out-of-plane shift of magnitude\cite{baskin_1955} $h=3.35\angstrom$. The Kek-Y deformations of the two layers are accounted for by the modulated hopping terms\cite{gamayun} (see Supplement)
\begin{equation}\label{eq:KekMonolayers}
\begin{split}
	t_{\rm B}(\rr,\ddelta_j)=&-\gamma_0[1+2\Delta_0\cos(\GG\cdot\rr)],\\
	t_{\rm T}(\rr,\ddelta_j)=&-\gamma_0[1+2\Delta_0\cos(\GG\cdot[\rr+\ddelta_j])],
\end{split}
\end{equation}
where $\gamma_0=3.16\,{\rm eV}$ is the intralayer hopping energy, $\Delta_0$ is the bond strength modulation due to the Kek-Y distorsion, $\GG=\KK_+-\KK_-$ is a Kekul\'e Bragg vector, and $\KK_{\pm}=\tfrac{4\pi}{3a}(\pm\tfrac{1}{2},\tfrac{\sqrt{3}}{2})$ are inequivalent graphene Brillouin zone (GBZ) corners (Fig.\ \ref{fig:main}a).

Based on the Slonczewski-Weiss-McClure model for graphite\cite{McClure,SlonWeiss}, we describe interlayer bonding with the Hamiltonian\footnote{Here, we ignore the interlayer intra-sublattice hopping term $\gamma_4$, which merely introduces particle-hole asymmetry\cite{mccann_falko}. We show in the Supplement that all our results remain valid when that term is included.}
\begin{equation}
\begin{split}
	H_1=\sum_{\RR}\Big[& \sum_{j,m=1}^3\tfrac{1-\delta_{j,m}}{2}t_{3}(\RR-\ddelta_j,\ddelta_j+\ddelta_m)c_{\RR-\ddelta_j}^\dagger b_{\RR+\ddelta_m}\\ &+ \gamma_1 a_{\RR}^\dagger d_{\RR} + \text{H.c.} \Big]
	%&+ \sum_{j=1}^3t_3(\RR+2\ddelta_j,-\ddelta_j)c_{\RR+2\ddelta_j}^\dagger b_{\RR+\ddelta_j}   \Big]
\end{split}
\end{equation}
Here, $\gamma_1=0.381\,{\rm eV}$ describes interlayer hopping between Kek-Y centers in the two layers, which is constant throughout the superlattice. By contrast, terms of the form (Supplement)
\begin{equation}\label{eq:t3}
	t_3(\rr,\ddelta)=-\gamma_3\left[1-2\Delta_3\cos{([2\KK_++\KK_-]\cdot\ddelta+\GG\cdot\rr)} \right],
\end{equation}
with $\gamma_3=0.38\,{\rm eV}$, produce the interlayer bond modulation shown in Fig.\ \ref{fig:main}b. Projected onto the plane, the pattern resembles an O-shaped Kekul\'e distortion (Kek-O, \cite{gamayun,Chamon2000}), geometrically analogous to what occurs naturally in benzene rings. Although Kek-O patterns have not been observed in graphene, but only in artificial graphene-like systems\cite{manoharan}, here, a Kek-O superlattice arises naturally for the $\gamma_3$ bonds as the separation between top-layer $A$- and bottom-layer $B$-sublattice atoms are modified by the in-plane Kekul\'e distortions. This is described by $\Delta_3$ in Eq.\ \eqref{eq:t3}. Considering in- and out-of-plane Gr\"uneisen parameters $\beta_{\parallel}$ and $\beta_{\perp}$, respectively, where\cite{Sun_2018,AMORIM20161,naumisReview} $\beta_{\perp}\approx 1.35 \beta_{\parallel} \approx 2.26$, we find 
\begin{equation}
	\Delta_3=\frac{2\beta_\perp}{\beta_\parallel}\left[1+\frac{3h^2}{a^2}\right]^{-1}\Delta_0.
\end{equation}

Fourier-transforming the full Hamiltonian $H=H_0+H_1$, and folding it into the Kekul\'e Brillouin zone (KBZ, Fig.\ \ref{fig:main}a), we obtain
\begin{equation}\label{eq:Hfull}
	H=\sum_{\kk \in {\rm KBZ}}\begin{pmatrix} \psi^\dagger_{\kk},&\chi^\dagger_{\kk} \end{pmatrix}\begin{pmatrix}
	h_\psi(\kk) & t(\kk)\\
	t^\dagger(\kk) & h_{\chi}(\kk)
	\end{pmatrix}\begin{pmatrix}\psi_{\kk} \\ \chi_{\kk} \end{pmatrix},
\end{equation}
where $\psi_{\kk}=(c_{\kk-\GG},\,b_{\kk-\GG},\,c_{\kk+\GG},\,b_{\kk+\GG})^T$ and $\chi_{\kk}=(a_{\kk+\GG},\,d_{\kk+\GG},\,a_{\kk-\GG},\,d_{\kk-\GG})^T$ are four-spinors; $\alpha_{\kk}=N^{-1/2}\sum_{\rr}e^{i\kk\cdot\rr}\alpha_{\rr}$ ($\alpha=a,\,b,\,c,\,d$), with $N$ the number of unit cells; and
\begin{widetext}
\begin{equation}
\begin{split}
	&h_\psi(\kk)=-\gamma_3\begin{pmatrix}
	0                                         & s_{-1}^*      & 0                                        & -\Delta_3s_0^*\\
	s_{-1}        & 0                                           & -\Delta_3s_{0}  & 0                                         \\
	0                                         & -\Delta_3s_{0}^*  & 0                                        & s_{1}^*    \\
	-\Delta_3s_0 & 0                                          & s_{1}                    & 0
	\end{pmatrix},\,
	h_\chi(\kk)=\gamma_1\begin{pmatrix}
	0                                         & 1                            & 0                                        & 0                                        \\
	1                          & 0                                           & 0                                        & 0                                         \\
	0                                         & 0                                           & 0                                        & 1                          \\
	0                                         & 0                                          & 1                          & 0
	\end{pmatrix},\, t(\kk) = -\gamma_0\begin{pmatrix}
	0                                                      & \Delta_0s_{-1} & 0                                                   & s_{-1}            \\
	\Delta_0s_{-1}^*     & 0                                               & s_{-1}^*               & 0                                             \\
	0                                                      & s_{1}             & 0                                                    &\Delta_0s_{1}\\
	s_{1}^*                 & 0                                               & \Delta_0s_{1}^*  & 0
	\end{pmatrix}.
\end{split}
\end{equation}
\end{widetext}
We have defined the functions $s_{\mu}(\kk)\equiv\sum_{j=1}^3e^{i(\kk+\mu\GG)\cdot\ddelta_j}$ ($\mu=-1,\,0,\,1$), and for simplicity, ignored all states near the GBZ $\Gamma$ point, which are $\sim 10\,{\rm eV}$ higher than the low-lying states, thus reducing the Hamiltonian's dimensions from $12\times12$ to $8\times8$. The $\chi$ subspace consists of interlayer dimers at energies $\pm \gamma_1$, which we may project out to obtain a $4\times4$ low-energy model\cite{mccann_falko}. As explained in the Supplement, we do so up to second order in perturbation theory by performing a standard L\"owdin partitioning\cite{lowdin}, and obtain the effective Bloch Hamiltonian about the $\Gamma$ point
\begin{equation}\label{eq:Heff}
\begin{split}
	H_{\rm eff}(\kk)=&-v(\sigma_1\tau_3k_x+\sigma_2\tau_0k_y)+(3\gamma_3\Delta_3 + 2w^2\Delta_0^2k^2)\sigma_1\tau_1\\
	 &-w^2(1+\Delta_0^2)\re \{ (k_x+ik_y)^2\}\sigma_1\tau_0\\
	 &+ w^2(1+\Delta_0^2)\im\{(k_x+ik_y)^2\}\sigma_2\tau_3 .
\end{split}
\end{equation}
We have defined $v=a\sqrt{3}\gamma_3/2$ and $w^2=3a^2\gamma_0^2/4\gamma_1$, and kept terms proportional to $\gamma_3$ up to first order in momentum. $\sigma_m$ and $\tau_n$ are Pauli matrices acting upon the sublattice and valley subspaces, respectively. Eq.\ \eqref{eq:Heff} is a good approximation to the full model \eqref{eq:Hfull}, as long as the Kekul\'e distortion is small ($\Delta_0 \lesssim 0.2$; see Fig.\ \ref{fig:main}c and Supplement).

Figs.\ \ref{fig:main}c--e show the main findings of this Letter: considering a small in-plane bond modulation of $\Delta_0\sim 0.1$, the low-energy dispersion of the $p6m$ Kek-Y graphene bilayer consists of a hexad of low-eccentricity elliptical Dirac bands. These bands can be divided into two sets, given by the Dirac points appearing at $\eta\kk_{\rm D} = \eta k_{\rm D}\hat{\vec{x}}$, with valley index $\eta=\pm 1$, and their corresponding symmetry partners obtained by $C_3$ rotations. We provide an analytic expression for $k_{\rm D}$, estimated from the effective model \eqref{eq:Heff}, in the Supplement. Here, we only point out that $k_{\rm D}$ is much smaller than the KBZ size ($k_{\rm D} \ll |\GG|/\sqrt{3}$).

The origin of the multi-valley structure can be traced to the Lifshitz transition occurring at either $K$ point in the band structure of Bernal-stacked bilayer graphene\cite{McCann2007}, where four Dirac cones appear below energies of order $1\,{\rm meV}$ (not shown; see, \emph{e.g.}, Ref.\ \onlinecite{Varlet_2015}). Folding these cones into the KBZ produces the hexagonal node structure, with a doubly-degenerate cone exactly at the KBZ center, shown in gray in the top inset of Fig.\ \ref{fig:main}c. It is clear, however, that the internode distance does not correspond to $k_{\rm D}$, which is instead determined by the Kekul\'e distortion strength, as described in the Supplement. More importantly, Fig.\ \ref{fig:main}c shows that the Dirac cone bandwidth in the $p6m$ Kekul\'e bilayer is about 30 times larger than the Lifsitz transition energy. This means that multi-valley physics should survive at room temperature, and for carrier densities as large as $7 \times 10^{11}{\rm cm}^{-2}$, or possibly higher if\cite{gutierrez} $\Delta_0>0.08$. By contrast, ultra-clean suspended or encapsulated samples are required to observe the Lifshitz transition in bilayer graphene\cite{Yacoby2010,Falko2014}. An important conclusion may be drawn by this connection; namely, that massless Dirac fermions appear only for Bernal-stacked Kekul\'e bilayers, where the Lifshitz transition is present, and not for $AA$ stacking, which we also classify in the Supplement.

%The degenerate central cones are also coupled by the distortion, and split into two pairs of parabolic electron and hole bands that touch exactly at the KBZ center. This central gap is about ($\sim 100\,{\rm meV}$), extending the linear regime of the six surrounding bands up to approximately $\pm25\,{\rm meV}$. That is, up to room temperature.

Linearizing $H_{\rm eff}(\kk)$ about $\eta \kk_{\rm D}$, and projecting out the higher-energy parabolic bands up to first order in momentum, we obtain
\begin{equation}\label{eq:H2by2}
	H_{\rm eff}(\eta \kk_{\rm D} + \qq) \approx \hbar\left(\eta V_xq_x \sigma_1 + V_yq_y\sigma_2\right),
\end{equation}
with expressions for the Fermi velocites $V_x,\,V_y$ given in the Supplement. For $\Delta_0=0.08$, we find $V_x=1.93\times 10^{-3}c$ and $V_y=1.23\times 10^{-3}c$, where $c$ is the speed of light in vacuum, corresponding to elliptical Dirac bands of eccentricity $0.77$. Similarly to the case of graphene, Eq.\ \eqref{eq:H2by2} gives opposite chiralities $\eta = \pm 1$ for states about the Dirac points located at $\eta C_3^n \kk_{\rm D}$ ($n=0,1,2$). This is verified numerically by the opposite Berry curvatures of adjacent valleys, evaluated using the full model \eqref{eq:Hfull}, and shown in Fig.\ \ref{fig:main}d. 

By contrast to graphene, however, the bilayer nature of the present system allows opening a relativistic mass gap when an out-of-plane electric field is applied (Fig.\ \ref{fig:main}e), producing a potential drop $\Delta V$ across the structure. This lowers the symmetry from $p6m$ to $p3$ by distinguishing between second-nearest interlayer neighbors, whose 2D projections would have otherwise been connected by $C_6$ rotations. For $|e\Delta V| \lesssim 50\,{\rm meV}$, the parameters of the effective model \eqref{eq:Heff} become slightly renormalized, and a new term,
\begin{equation}\label{eq:DVeff}
    \Delta V_{\rm eff} = \frac{e\Delta V}{2}\left[1-\frac{\left(1- \frac{\gamma_4^2}{\gamma_0^2}\right)\left(1+\Delta_0^2\right) }{\gamma_1+\frac{e^2\Delta V^2}{4\gamma_1}}w^2k^2\right]\sigma_3\tau_0,
\end{equation}
is introduced, where $e$ is the (positive) electronic charge, and $v_{0,4}=a\gamma_{0,4}\sqrt{3}/2$. Similar effects (gap opening and symmetry lowering) are produced by intralayer sublattice asymmetries. By contrast, both the $p6m$ symmetry and the gapless Dirac spectrum are preserved by an on-site energy asymmetry between dimerizing- and non-dimerizing sites, caused by different molecular environments\cite{Marcin2010}.

\begin{figure}[t!]
    \centering
    \includegraphics[width=1.0\columnwidth]{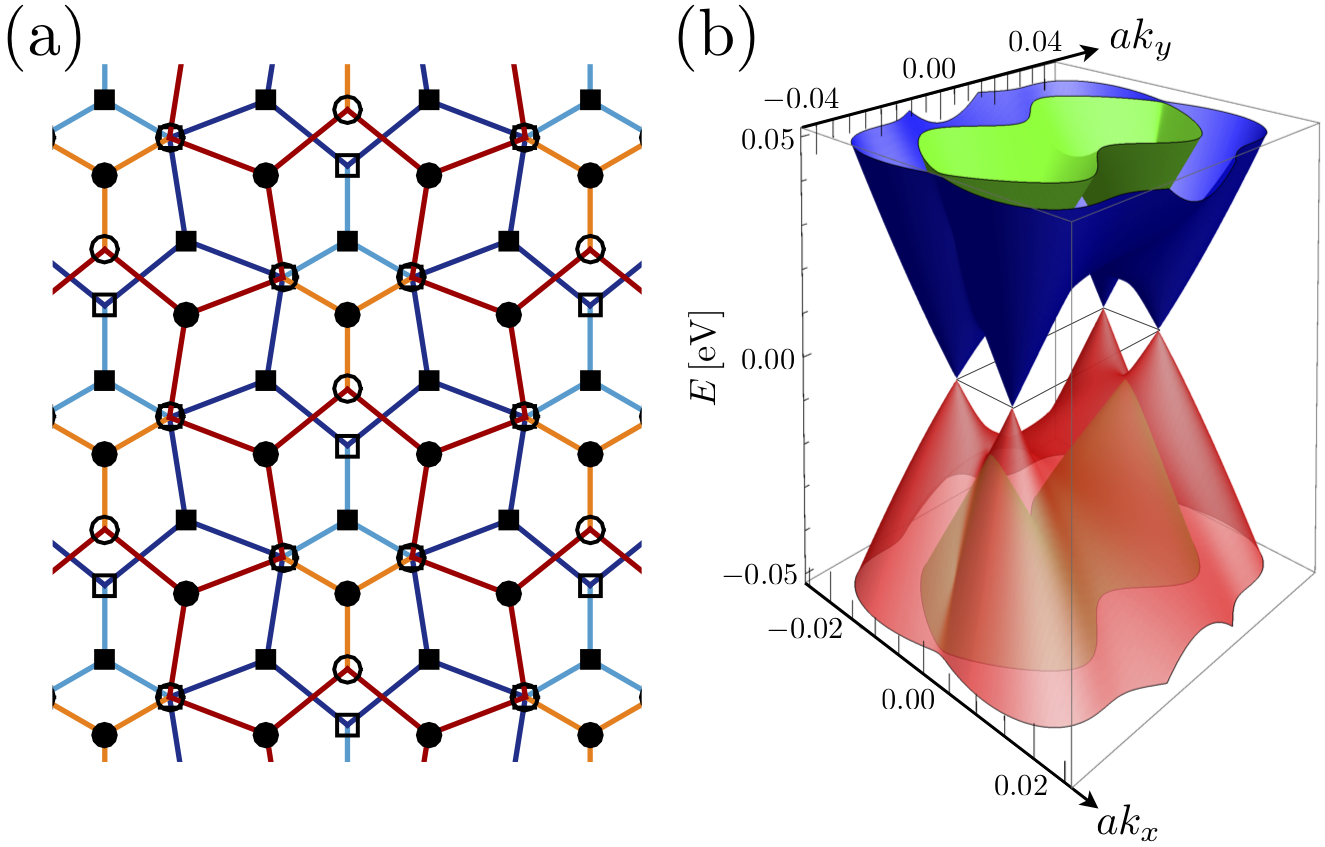}\label{fig:categories}
    \caption{(a) Bilayer obtained by the method of Fig.\ \ref{fig:device}, where the Kekul\'e deformation centers are separated along the $\hat{\mathbf{y}}$ direction. As a result, chains of intra- and interlayer short bonds distribute preferentially along this axis. (b) Corresponding multi-valley band structure, evaluated with $\Delta_0=0.1$.}
    \label{fig:cmm}
\end{figure}

 So far, we have focused on the most symmetric configuration that can result from stacking the two Kekul\'e-distorted graphene layers. A different arrangement, described by the space group $cmm$, is shown in Fig.\ \ref{fig:cmm}a, and its corresponding low-energy dispersion in Fig.\ \ref{fig:cmm}b. By performing numerical tight-binding calculations, we found that this system exhibits 4 inequivalent Dirac cones distributed preferentially along the $\hat{\vec{y}}$ axis, mirroring the formation of chains of distorted interlayer bonds in its structure (see Fig.\ \ref{fig:cmm}a). Other possible patterns are shown in the Supplement, including a chiral structure described by the $p6$ wallpaper group, for which we anticipate interesting optical activity. An in-depth exploration of the symmetry and electronic properties of this large family of materials will be published elsewhere\cite{us2020}.
 
% while we present all $8$ patterns. The space group determines the resulting  quantum phase. For example, in Figure ** we present the energy dispersion for the $**$ pattern. Instead of six, two Dirac cones are observed. Other patterns present one-dimensional interacting chains, which in principle suggests phases close to a Luttinger liquid. The patterns with $p6$ wallpapergroup are interesting as they  present chirality and thus are expected to display interesting optical activity.

Multi-valley systems host vastly complex phenomena, both from the single- and many-particle perspectives. Here, we have proposed a family of Bernal-stacked bilayer graphene Kekul\'e superlattices that display Dirac bands centered at multiple valleys near the mini Brillouin zone center. In the most symmetric case, where the Kekul\'e distortion centers of the two layers coincide, we have predicted a system of multi-valley Dirac fermions with electrostatically tunable mass gaps, with great potential for many-body and quantum Hall physics at room temperature. The small momentum-space separation ($\sim 10^{-2}\,\angstrom^{-1}$) between Dirac valleys in all cases studied makes the band structure susceptible to long-range disorder, requiring samples with high crystalline quality to explore their multi-valley spectra. If Coulomb interactions are considered, the Fermi velocities $V_{x,y}$ reported above give an effective fine structure constant of $\alpha_{\rm eff}=e^2/\hbar V_{x,y} \approx 4.6$, corresponding to strong coupling\cite{semenoff2012}. This makes the different Kekul\'e graphene bilayers studied in this Letter good candidates to observe many-body effects, including the anomalous quantum Hall insulating\cite{Levitov2010} and gapless symmetry-broken\cite{lemonik2010,Vafek2010} phases predicted for bilayer graphene. We believe that systems like the one proposed in this Letter may just be within the reach of current experimental capabilities, drawing from state-of-the art techniques currently used for manual stacking of van der Waals crystals. We anticipate that uncertainties in the exact interlayer alignment, inherent in the type of systems we propose, will lead to moir\'e pattern formation, adding a further layer of complexity to the already rich phenomenology predicted in this Letter.

%\emph{Conclusions}\label{sec:conclusions}
%Multi-valley systems host vastly complex phenomena, both from the single- and many-particle perspectives. SU(6) symmetry has been reported in Bismuth (111) surface states\cite{Bi2001,Bi2016}, and, recently, a prediction for SU(3) quark physics with spontaneous symmetry breaking in $n$-doped transition-metal dichalcogenides has been put forth\cite{TMD_SU3}. Here, we have proposed a family of bilayer graphene Kekul\'e superlattices that display Dirac bands centered at multiple valleys. In the most symmetric case of Bernal stacking, with coinciding distortion centers in both layers, we have predicted a system of strongly interacting SU(2)$\times$SU(6) Dirac fermions, with great potential for many-body and quantum Hall physics near room temperature. Devices like the one proposed in this Letter may just be within the reach of current experimental capabilities, drawing from state-of-the art techniques currently used for manual stacking of van der Waals crystals.

D.R.-T.\ acknowledges funding from UNAM-DGAPA.  E.A.\ thanks CONACyT for a doctoral scholarship. R.C.-B. thanks 20va Convocatoria Interna (UABC). E.A.\ and G.N.\ thank UNAM-DGAPA-PAPIIT project IN-102717 for finantial support. F.M.\ thanks the support of PAPIIT-UNAM, through project IN111317. D.R.-T.\ thanks M.\ Asmar for useful discussions during preparation of this paper.

\bibliographystyle{apsrev4-1}
\bibliography{biblio.bib}
\end{document}